\documentclass[twocolumn,showkeys,showpacs,preprintnumbers,amsmath,amssymb]{revtex4}

\usepackage{graphicx}
\usepackage{dcolumn}
\usepackage{bm}
\usepackage{subfigure}

\begin{document}

\preprint{APS/123-QED}

\title{State Space Reconstruction for Multivariate Time Series Prediction}

\author{I. Vlachos}
 \email{ivlaxos@gen.auth.gr}
\author{D. Kugiumtzis}%
 \email{dkugiu@gen.auth.gr}
\affiliation{Department of Mathematical, Physical and
Computational Sciences, Faculty of Technology, Aristotle
University of Thessaloniki, Greece}

\date{\today}

\begin{abstract}
In the nonlinear prediction of scalar time series, the common practice is to reconstruct the state space using
time-delay embedding and apply a local model on neighborhoods of the reconstructed space. The method of false nearest
neighbors is often used to estimate the embedding dimension. For prediction purposes, the optimal embedding dimension
can also be estimated by some prediction error minimization criterion. We investigate the proper state space
reconstruction for multivariate time series and modify the two abovementioned criteria to search for optimal embedding
in the set of the variables and their delays. We pinpoint the problems that can arise in each case and compare the
state space reconstructions (suggested by each of the two methods) on the predictive ability of the local model that
uses each of them. Results obtained from Monte Carlo simulations on known chaotic maps revealed the non-uniqueness of
optimum reconstruction in the multivariate case and showed that prediction criteria perform better when the task is
prediction.
\end{abstract}

 \maketitle

\section{Introduction}

Since its publication \emph{Takens' Embedding Theorem} \cite{Takens81} (and its extension, the \emph{Fractal Delay
Embedding Prevalence Theorem} by Sauer et al. \cite{Sauer91}) has been used in time series analysis in many different
settings ranging from system characterization and approximation of invariant quantities, such as correlation dimension
and Lyapunov exponents, to prediction and noise-filtering \cite{Kantz97}. The Embedding Theorem implies that although
the true dynamics of a system may not be known, equivalent dynamics can be obtained under suitable conditions using
time delays of a single time series, treated as an one-dimensional projection of the system trajectory.

Most applications of the Embedding Theorem deal with univariate time series, but often measurements of more than one
quantities related to the same dynamical system are available. One of the first uses of multivariate embedding was in
the context of spatially extended systems where embedding vectors were constructed from data representing the same
quantity measured simultaneously at different locations \cite{Guckenheimer83,Palus92}. Multivariate embedding was used
for noise reduction \cite{Hegger92} and for surrogate data generation with equal individual delay times and equal
embedding dimensions for each time series \cite{Prichard94}. In nonlinear multivariate prediction, the prediction with
local models on a space reconstructed from a different time series of the same system was studied in
\cite{Abarbanel94}. This study was extended in \cite{Cao98} by having the reconstruction utilize all of the observed
time series. Multivariate embedding with the use of independent components analysis was considered in \cite{Barnard01}
and more recently multivariate embedding with varying delay times was studied in \cite{garcia05,hirata06}.

In this work, we focus on the state space reconstruction from multivariate time series from discrete dynamical systems,
so that the investigation of optimal embedding does not involve the delay parameter but only the embedding dimension
for each variable. For this, we adjust two well-known approaches used for univariate time series, i.e. the false
nearest neighbor \cite{Kennel92} and the optimal reconstruction for prediction evaluated in a test set
\cite{kaplan93,Chun-Hua04}. We study the consistency of the techniques in estimating the embedding dimensions as well
as their performance by means of out-of-sample prediction using the selected embedding. Monte Carlo simulations at
different settings of system complexity, system dimension and time series lengths are used to evaluate the embedding
techniques.

In Section~\ref{sec:uniembedding} the embedding for univariate time series is briefly discussed. In
Section~\ref{sec:multiembedding} the discussion is extended to multivariate time series and the suggested techniques
for estimating the embedding are presented. Then in Section~\ref{sec:MonteCarlo} the results of Monte Carlo simulations
are presented and in Section~\ref{sec:Discussion} the results are discussed and conclusions are given.

\section{Univariate Embedding}
\label{sec:uniembedding}

A dynamical system generates a trajectory in a $D$-dimensional manifold $\Gamma$. For discrete time the dynamical
system is defined by the $D$-dimensional map $\mathbf{F}:\Gamma \mapsto \Gamma$ as
\[
\mathbf{y}_{n+1}=\mathbf{F}(\mathbf{y}_{n}), \quad
 n \in \mathbb{N},
\]
where $\mathbf{y}_n \in \Gamma$ is the state vector at time
step $n$.

The observed scalar time series $\{x_n\}_{n=1}^N$ of length $N$ is the projection of the segment of the system
trajectory $\{\mathbf{y}_n\}_{n=1}^N$ given by a measurement function $h: \Gamma \mapsto \mathbb{R}$ as $
x_n=h(\mathbf{y}_n)$. Despite the apparent loss of information of the system dynamics by the projection, the system
dynamics may be recovered through suitable state space reconstruction from the scalar time series.

\subsection{Reconstruction of the state space}

According to Taken's embedding theorem a trajectory formed by the points $\mathbf{x}_n$ of time-delayed components from
the time series $\{x_n\}_{n=1}^N$ as
\begin{equation}
\label{embvec}
\mathbf{x}_n=(x_{n-(m-1)\tau},x_{n-(m-2)\tau},...,x_n),
\end{equation}
under certain genericity assumptions, is an one-to-one mapping of the original trajectory of $\mathbf{y}_n$ provided
that $m$ is large enough.

Given that the dynamical system ``lives'' on an attractor $A \subset \Gamma$, the reconstructed attractor $\tilde{A}$
through the use of the time-delay vectors is topologically equivalent to $A$. A sufficient condition for an appropriate
unfolding of the attractor is $m \ge 2d+1$ where $d$ is the box-counting dimension of $A$.

The embedding process is visualized in the following graph
\begin{center}
\begin{tabular}
[c]{cclcc}%
$\mathbf{y}_{n}\in A$ & $\subset$ & $\Gamma$ & $\quad
\overset{\mathbf{F}}{\mapsto}$ &
$\quad \mathbf{y}_{n+1}\in A\subset\Gamma$\\
$\downarrow_{h}$ & & & & $\quad \downarrow_{h}$\\
$x_{n}\in\mathbb{R}$ & & & & $\quad x_{n+1}\in\mathbb{R}$\\
$\downarrow_{e}$ & & & & $\quad \downarrow_{e}$\\
$\mathbf{x}_{n}\in\tilde{A}$ & $\subset$ & $\mathbb{R}^{m}$ &
$\quad \overset{\mathbf{G}}{\mapsto}$ &
$\quad \mathbf{x}_{n+1}\in\tilde{A}\subset\mathbb{R}^{m}$%
\end{tabular}
\end{center}
where $e$ is the embedding procedure creating the delay vectors from the time series and $\mathbf{G}$ is the
reconstructed dynamical system on $\tilde{A}$. $\mathbf{G}$ preserves properties of the unknown $\mathbf{F}$ on the
unknown attractor $A$ that do not change under smooth coordinate transformations.

\subsection{Univariate local prediction}

For a given state space reconstruction, the local prediction at a target point $\mathbf{x}_n$ is made with a model
estimated on the $K$ nearest neighboring points to $\mathbf{x}_n$. The local model can have a simple form, such as the
zeroth order model (the average of the images of the nearest neighbors), but here we consider the linear model
\begin{equation}
\label{locmod} \nonumber
\hat{x}_{n+1}=\mathbf{a}^{(n)}\mathbf{x}_n+b^{(n)},
\end{equation}
where the superscript $(n)$ denotes the dependence of the model parameters ($\mathbf{a}^{(n)}$ and $b^{(n)}$) on the
neighborhood of $\mathbf{x}_n$. The neighborhood at each target point is defined either by a fixed number $K$ of
nearest neighbors or by a distance determining the borders of the neighborhood giving a varying $K$ with
$\mathbf{x}_n$.

\subsection{Selection of embedding parameters}

The two parameters of the delay embedding in \eqref{embvec} are the embedding dimension $m$, i.e. the number of
components in $\mathbf{x}_n$ and the delay time $\tau$. We skip the discussion on the selection of $\tau$ as it is
typically set to $1$ in the case of discrete systems that we focus on. Among the approaches for the selection of $m$ we
choose the most popular method of {\em false nearest neighbors} (FNN) and present it briefly below \cite{Kennel92}.

The measurement function $h$ projects distant points $\{\mathbf{y}_n\}$ of the original attractor to close values of
$\{x_n\}$. A small $m$ may still give badly projected points and we seek the reconstructed state space of the smallest
embedding dimension $m$ that unfolds the attractor. This idea is implemented as follows. For each point
$\mathbf{x}_n^m$ in the $m$-dimensional reconstructed state space, the distance from its nearest neighbor
$\mathbf{x}_{n(1)}^m$ is calculated, $d(\mathbf{x}_n^m,\mathbf{x}_{n(1)}^m) =
\|{\mathbf{x}}_n^m-{\mathbf{x}}_{n(1)}^m\|$. The dimension of the reconstructed state space is augmented by 1 and the
new distance of these vectors is calculated, $d(\mathbf{x}_n^{m+1},\mathbf{x}_{n(1)}^{m+1})
=\|\mathbf{x}_n^{m+1}-\mathbf{x}_{n(1)}^{m+1}\|$. If the ratio of the two distances exceeds a predefined tolerance
threshold $r$ the two neighbors are classified as false neighbors, i.e.
\begin{equation}
r_n(m)=\frac{d(\mathbf{x}_n^{m+1},\mathbf{x}_{n(1)}^{m+1})}{d(\mathbf{x}_n^m,\mathbf{x}_{n(1)}^m)}
> r.
\label{eq:FNNcheck}
\end{equation}

The criterion that the embedding dimension $m$ is high enough to unfold the attractor is that the percentage of points
for which $r_n(m)>r$, is essentially zero, typically requiring to be smaller than $1\%$.

The selection of $r$ should be large enough to allow for exponential divergence. In \cite{Kantz99}, a stricter
criterion is introduced, that the original distance of the point to its nearest neighbor in the $m$-dimensional space
does not exceed the standard deviation of $x_n$ divided by $r$. If it does the point is omitted from the percentage
calculation, since the points are already too far apart to be real neighbors. A good and often used value for $r$ is
10.

Another popular method for the selection of the embedding dimension $m$ is from the optimization of the fit of a local
linear model using a criterion for the goodness-of-fit \cite{kaplan93} or the goodness-of-prediction \cite{Chun-Hua04}.
The idea here is that for a local linear model fit to be optimum the attractor must be fully unfolded. After the
selection of an appropriate $\tau$, for state space reconstructions with $m$ varying from 1 to a maximum
$m_{\mathrm{max}}$, the fit or prediction error of local prediction models is calculated. For the errors a statistic
such as the {\em normalized root mean square error} (NRMSE) is used and the embedding dimension is chosen as the one
that minimizes this statistic. Whereas the false nearest neighbors method determines a minimal sufficient embedding
dimension, this method picks a dimension for which the attractor is unfolded (so as to give better predictions) that
may be larger than the minimal.

\section{Multivariate Embedding}
\label{sec:multiembedding}

In Section~\ref{sec:uniembedding} we gave a summary of the reconstruction technique for a deterministic dynamical
system from a scalar time series generated by the system. However, it is possible that more than one time series are
observed that are possibly related to the system under investigation. For $p$ time series measured simultaneously from
the same dynamical system, a measurement function $\mathbf{H}:\Gamma\mapsto\mathbb{R}^p$ is decomposed to $h_i$,
$i=1,\ldots,p$, defined as in Section~\ref{sec:uniembedding}, giving each a time series $\{x_{i,n}\}_{n=1}^N$.
According to the discussion on univariate embedding any of the $p$ time series can be used for reconstruction of the
system dynamics, or better, the most suitable time series could be selected after proper investigation. In a different
approach all the available time series are considered and the analysis of the univariate time series is adjusted to the
multivariate time series.

\subsection{From univariate to multivariate embedding}
\label{sec:scamul}

Given that there are $p$ time series $\{x_{i,n}\}_{n=1}^N$, $i=1,\ldots,p$, the equivalent to the reconstructed state
vector in \eqref{embvec} for the case of multivariate embedding is of the form
\begin{equation}
 \begin{split}
 \label{multembvec}
 \mathbf{x}_n &=(x_{1,n-(m_1-1)\tau_1},x_{1,n-(m_1-2)\tau_1},...,x_{1,n},\\
&\quad\quad\quad\quad\quad\quad\quad
x_{2,n-(m_2-1)\tau_2},...,x_{2,n},...,x_{p,n})
 \end{split}
 \end{equation}
and are defined by an \emph{embedding dimension vector} $\mathbf{m}=(m_{1},...,m_{p})$ that indicates the number of
components used from each time series and a \emph{time delay vector} ${\pmb\tau}=(\tau_{1},...,\tau_{p})$ that gives
the delays for each time series. The corresponding graph for the multivariate embedding process is shown below.
\begin{center}
\begin{tabular}
[c]{ccc}
$\mathbf{y}_{n}\in A$ \,\, $\subset$ \, $\Gamma$ & $\quad \overset{\mathbf{F}}{\mapsto}$%
& $\mathbf{y}_{n+1}\in A\subset\Gamma$\\%
\begin{tabular}
[c]{lll}%
$\swarrow_{h_{1}}$ & $\downarrow_{h_{2}} ...$ & $\searrow_{h_{p}}$%
\end{tabular}
 &  &
\begin{tabular}
[c]{lll}%
$\swarrow_{h_{1}}$ & $\downarrow_{h_{2}} ...$ & $\searrow_{h_{p}}$%
\end{tabular}
\\%
\begin{tabular}
[c]{lll}%
$x_{1,n}$\ \ \  & $x_{2,n} ...$\ \ \  & $x_{p,n}$%
\end{tabular}
 &  &
\begin{tabular}
[c]{lll}%
$x_{1,n+1}$ & $x_{2,n+1} ...$ & $x_{p,n+1}$%
\end{tabular}
\\%
\begin{tabular}
[c]{lll}%
$\searrow_{e}$\ \  & $\downarrow_{e} ...$\ \  & $\swarrow_{e}$%
\end{tabular}
 &  &
\begin{tabular}
[c]{lll}%
$\searrow_{e}$\ \  & $\downarrow_{e} ...$\ \  & $\swarrow_{e}$%
\end{tabular}
\\
$\mathbf{x}_{n}\in\tilde{A}$ \,\, $\subset$ \, $\mathbb{R}^{M}$ & $\quad \overset{\mathbf{G}%
}{\mapsto}$ & $\mathbf{x}_{n+1}\in\tilde{A}\subset\mathbb{R}^{M}$\\
\end{tabular}
\end{center}
The \emph{total embedding dimension} $M$ is the sum of the individual embedding dimensions for each time series
$M=\sum_{i=1}^{p}m_i$. Note that if redundant or irrelevant information is present in the $p$ time series, only a
subset of them may be represented in the optimal reconstructed points $\mathbf{x}_n$. The selection of $\mathbf{m}$ and
${\pmb\tau}$ follows the same principles as for the univariate case: the attractor should be fully unfolded and the
components of the embedding vectors should be uncorrelated. A simple selection rule suggests that all individual delay
times and embedding dimensions are the same, i.e. $\mathbf{m}=m \mathbf{1}$ and ${\pmb\tau}=\tau\mathbf{1}$ with
$\mathbf{1}$ a $p$-vector of ones \cite{Hegger92,Prichard94}. Here, we set again $\tau_i=1$, $i=1,\ldots,p$, but we
consider both fixed and varying $m_i$ in the implementation of the FNN method (see Section~\ref{subsec:multiFNN}).

\subsection{Multivariate local prediction}

The prediction for each time series $x_{i,n}$, $i=1,\ldots,p$, is performed separately by $p$ local models, estimated
as in the case of univariate time series, but for reconstructed points formed potentially from all $p$ time series as
given in \eqref{multembvec} (e.g. see \cite{Cao98}).



We propose an extension of the NRMSE for the prediction of one time series to account for the error vectors comprised
of the individual prediction errors for each of the predicted time series. If we have one step ahead predictions for
the $p$ available time series, i.e. $\hat{x}_{i,n}$, $i=1,\ldots,p$ (for a range of current times $n-1$), we define the
multivariate NRMSE
\begin{equation}
\mbox{NRMSE}=\sqrt{{\sum\limits_{n}\|(x_{1,n}-\hat{x}_{1,n},\ldots,x_{p,n}-\hat{x}_{p,n})\|^{2} \over \sum\limits_{n}\|
(x_{1,n}-\bar{x}_{1},\ldots,x_{p,n}-\bar{x}_{p})\|^{2}}} \label{eq:multiNRMSE}
\end{equation}
where $\bar{x}_{i}$ is the mean of the actual values of $x_{i,n}$ over all target times $n$.

\subsection{Problems and restrictions of multivariate reconstructions}

A major problem in the multivariate case is the \emph{problem of identification}. There are often not unique
$\mathbf{m}$ and ${\pmb\tau}$ embedding parameters that unfold fully the attractor. A trivial example is the Henon map
\cite{Henon76}
\begin{equation}
 \begin{array}{l}
 x_{n+1}=1.4-x_{n}^2+y_{n} \\
 y_{n+1}=0.3x_{n}
 \end{array}
\label{eq:Henon}
\end{equation}
It is known that for the state space reconstruction from the observable $x_n$ the appropriate embedding parameters are
$m=2$ and $\tau=1$. Due to the fact that $y_n$ is a lagged multiple of $x_n$ the attractor can obviously be
reconstructed from the bivariate time series $\{x_n,y_n\}$ equally well with any of the following two-dimensional
embedding schemes
\[
 \mathbf{x}_n=(x_n,x_{n-1}) \quad \mathbf{x}_n=(x_n,y_n) \quad
 \mathbf{x}_n=(y_n,y_{n-1})
\]
since they are essentially the same. This example shows also the problem of {\em redundant information}, e.g. the state
space reconstruction would not improve by augmenting the delay vector $\mathbf{x}_n=(x_n,x_{n-1})$ with the component
$y_n$ that actually duplicates $x_{n-1}$. Redundancy is inevitable in multivariate time series as synchronous
observations of the different time series are generally correlated and the fact that these observations are used as
components in the same embedding vector adds redundant information in them. We note here that in the case of continuous
dynamical systems, the delay parameter ${\tau_i}$ may be selected so that the components of the $i$ time series are not
correlated with each other, but this does not imply that they are not correlated to components from another time
series.

A different problem is that of {\em irrelevance}, when time series that are not generated by the same dynamical system
are included in the reconstruction procedure. This may be the case even when a time series is connected to a time
series generated by the system under investigation.

An issue of concern is also the fact that multivariate data don't always have the same data ranges and distances
calculated on delay vectors with components of different ranges may depend highly on only some of the components. So it
is often preferred to scale all the data to have either the same variance or be in the same data range. For our study
we choose to scale the data to the range $[0,1]$.

\subsection{Selection of the embedding dimension vector}
\label{subsec:multiFNN}

Taking into account the problems in the state space reconstruction from multivariate time series, we present three
methods for determining $\mathbf{m}$, two based on the false nearest neighbor algorithm, which we name FNN1 and FNN2,
and one based on local models which we call \emph{prediction error minimization} criterion (PEM).

The main idea of the FNN algorithms is as for the univariate case. Starting from a small value the embedding dimension
is increased by including delay components from the $p$ time series and the percentage of the false nearest neighbors
is calculated until it falls to the zero level. The difference of the two FNN methods is on the way that $\mathbf{m}$
is increased.

\begin{figure}[t]
 \vspace{5mm}
 \subfigure[] {
\includegraphics[scale=0.25]{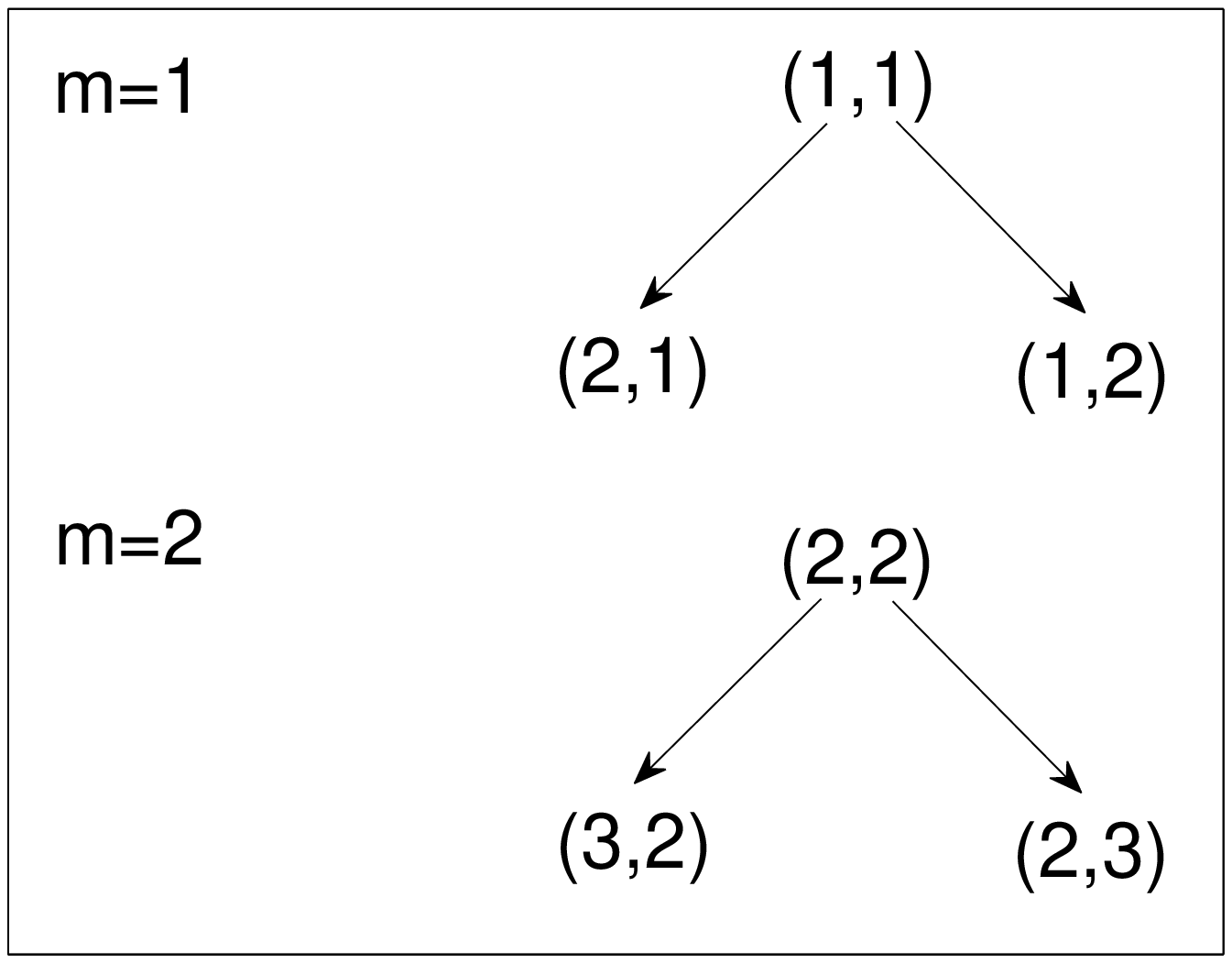}\label{fig:fnn:a}
}\hspace{2mm} \subfigure[] {
\includegraphics[scale=0.25]{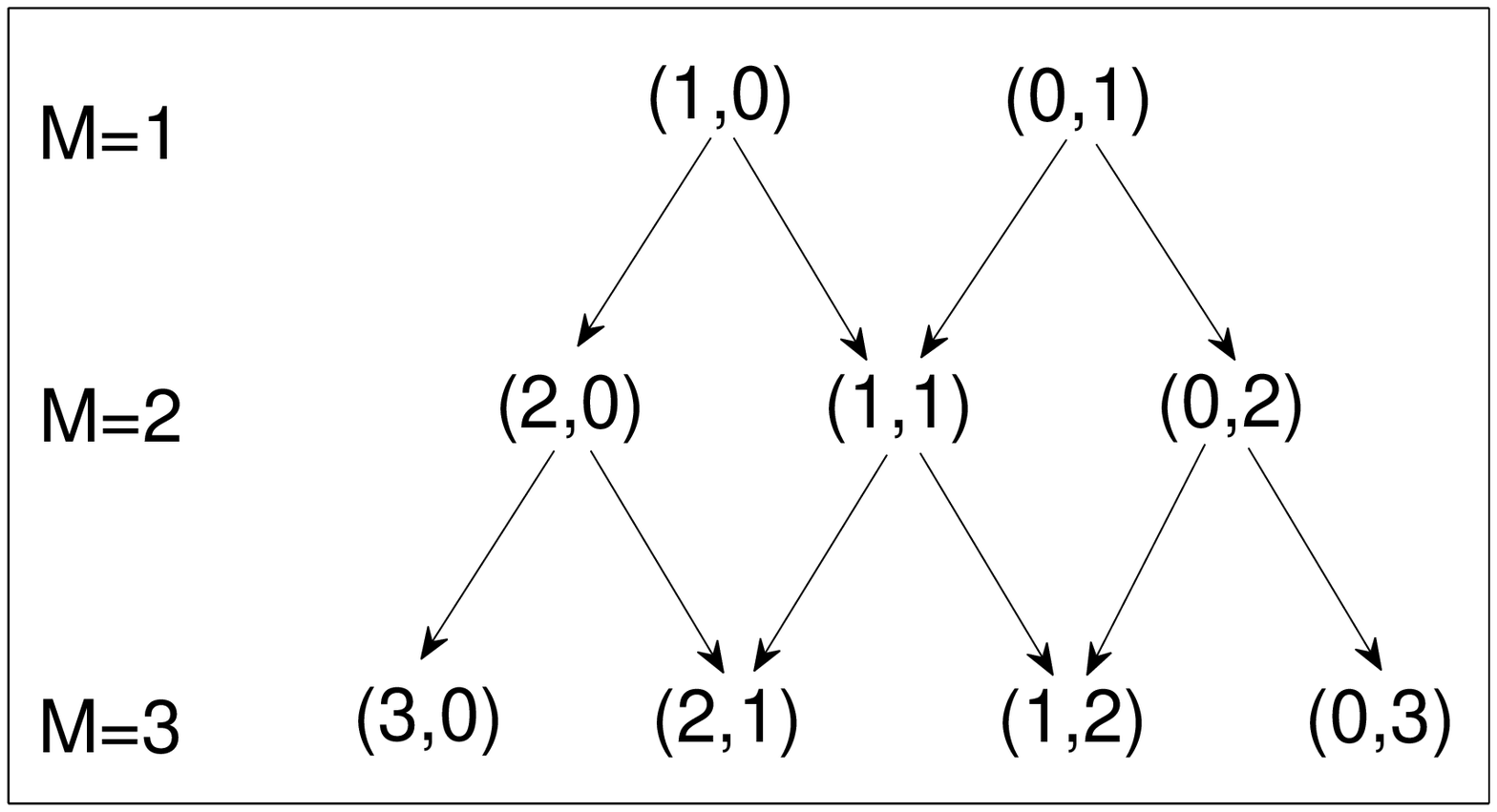}\label{fig:fnn:b}
} \caption{\label{fig:fnn} Example of the first two steps of FNN1 (a) and FNN2 (b) for a bivariate time series}
\end{figure}

For FNN1 we restrict the state space reconstruction to use the same embedding dimension for each of the $p$ time
series, i.e. $\mathbf{m}=(m,m,...,m)$ for a given $m$. To assess whether $\mathbf{m}$ is sufficient, we consider all
delay embeddings derived by augmenting the state vector of embedding dimension vector $(m,m,...,m)$ with a single
delayed variable from any of the $p$ time series. Thus the check for false nearest neighbors in \eqref{eq:FNNcheck}
yields the increase from the embedding dimension vector $(m,m,...,m)$ to each of the embedding dimension vectors
$(m+1,m,...,m)$, $(m,m+1,...,m)$, $\ldots$, $(m,m,...,m+1)$. Then the algorithm stops at the optimal
$\mathbf{m}=(m,m,...,m)$ if the zero level percentage of false nearest neighbors is obtained for all $p$ cases. A
sketch of the first two steps for a bivariate time series is shown in Figure~\ref{fig:fnn:a}.

This method has been commonly used in multivariate reconstruction and is more appropriate for spatiotemporally
distributed data (e.g. see the software package TISEAN \cite{Heggertisean99}). A potential drawback of FNN1 is that the
selected total embedding dimension $M$ is always a multiple of $p$, possibly introducing redundant information in the
embedding vectors.

We modify the algorithm of FNN1 to account for any form of the embedding dimension vector $\mathbf{m}$ and the total
embedding dimension $M$ is increased by one at each step of the algorithm. Let us suppose that the algorithm has
reached at some step the total embedding dimension $M$. For this $M$ all the combinations of the components of the
embedding dimension vector $\mathbf{m}=(m_1,m_2,...,m_p)$ are considered under the condition $M=\sum_{i=1}^{p}m_i$.
Then for each such $\mathbf{m}=(m_1,m_2,...,m_p)$ all the possible augmentations with one dimension are checked for
false nearest neighbors, i.e. $(m_1+1,m_2,...,m_p)$, $(m_1,m_2+1,...,m_p)$, $\ldots$, $(m_1,m_2,...,m_p+1)$. A sketch
of the first two steps of the extended FNN algorithm, denoted as FNN2, for a bivariate time series is shown in
Figure~\ref{fig:fnn:b}.
%

The termination criterion is the drop of the percentage of false nearest neighbors to the zero level at every increase
of $M$ by one for at least one embedding dimension vector $(m_1,m_2,...,m_p)$. If more than one embedding dimension
vectors fulfill this criterion, the one with the smallest cumulative FNN percentage is selected, where the cumulative
FNN percentage is the sum of the $p$ FNN percentages for the increase by one of the respective component of the
embedding dimension vector.

The PEM criterion for the selection of $\mathbf{m}=(m_1,m_2,...,m_p)$ is simply the extension of the goodness-of-fit or
prediction criterion in the univariate case to account for the multiple ways the delay vector can be formed from the
multivariate time series. Thus for all possible $p$-plets of $(m_1,m_2,...,m_p)$ from $(1,0,...,0)$, $(0,1,...,0) $,
etc up to some vector of maximum embedding dimensions $(m_{\mathrm{max}},m_{\mathrm{max}},\ldots,m_{\mathrm{max}})$,
the respective reconstructed state spaces are created, local linear models are applied and out-of-sample prediction
errors are computed. So, totally $p^{m_{\mathrm{max}}}-1$ embedding dimension vectors are compared and the optimal is
the one that gives the smallest multivariate NRMSE as defined in \eqref{eq:multiNRMSE}.

\section{Monte Carlo simulations and results}
\label{sec:MonteCarlo}

\subsection{Monte Carlo setup}

We test the three methods by performing Monte Carlo simulations on a variety of known nonlinear dynamical systems. The
embedding dimension vectors are selected using the three methods on 100 different realizations of each system and the
most frequently selected embedding dimension vectors for each method are tracked. Also, for each realization and
selected embedding dimension vector from each method, the multivariate NRMSE for out-of-sample prediction is computed.
The average multivariate NRMSE over the 100 realizations for each method is then used as an indicator of the
performance of each method in prediction.

\begin{figure}[t]
\includegraphics[scale=0.42]{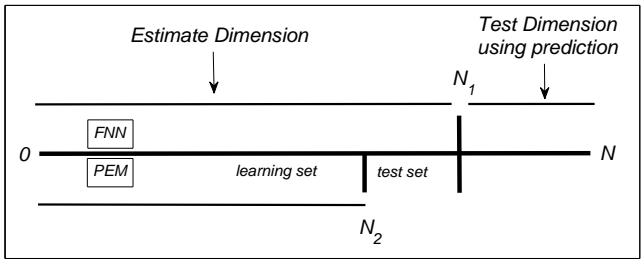}
\caption{\label{fig:dem2} Sketch of the split of data for the selection of the embedding dimension vector with FNN and
PEM.}
\end{figure}

The selection of the embedding dimension vector by FNN1, FNN2 and PEM is done on the first three quarters of the data,
$N_1=3N/4$, and the multivariate NRMSE is computed on the last quarter of the data ($N-N_1$). For PEM, the same split
is used on the $N_1$ data, so that $N_2=3N_1/4$ data are used to find the neighbors (training set) and the rest
$N_1-N_2$ are used to compute the multivariate NRMSE (test set) and decide for the optimal embedding dimension vector.
A sketch of the split of the data is shown in Figure \ref{fig:dem2}.
The number of neighbors for the local models in PEM varies with $N$ and we set $K_N=10,25,50$ for time series lengths
$N=512,2048,8192$, respectively. The parameters of the local linear model are estimated by ordinary least squares. For
all methods the investigation is restricted to $m_{\mathrm{max}}=5$.

The multivariate time series are derived from nonlinear maps of varying dimension and complexity as well as spatially
extended maps. The results are given below for each system.

\subsection{One and two Ikeda maps}

The Ikeda map is an example of a discrete low-dimensional chaotic system in two variables $(x_n,y_n)$ defined by the
equations \cite{Ikeda79}
\begin{eqnarray}
 z_{n+1}=1+0.9\exp({0.4{\mathrm i}-6{\mathrm i}/(1+|z_{n}|^2)}), \nonumber \\
 x_{n}=\mathrm{Re}(z_{n}), \quad\quad y_{n}=\mathrm{Im}(z_{n}),
 \nonumber
\end{eqnarray}
where $\mathrm{Re}$ and $\mathrm{Im}$ denote the real and imaginary part, respectively, of the complex variable $z_n$.
Given the bivariate time series of $(x_n,y_n)$, both FNN methods identify the original vector $\mathbf{x}_n=(x_n,y_n)$
and find $\mathbf{m}=(1,1)$ as optimal at all realizations, as shown in Table~\ref{tab:ikeda}.

On the other hand, the PEM criterion finds over-embedding as optimal, but this improves slightly the prediction, which
as expected improves with the increase of $N$.

Next we consider the sum of two Ikeda maps as a more complex and higher dimensional system. The bivariate time series
are generated as
\[
 x_n=\mathrm{Re}(z_{1,n}+z_{2,n}), \quad y_n=\mathrm{Im}(z_{1,n}+z_{2,n}).
\]
The results of the Monte Carlo simulations shown in Table~\ref{tab:ikeda2} suggest that the prediction worsens
dramatically from that in Table~\ref{tab:ikeda} and the total embedding dimension $M$ increases with $N$.
\begin{table}[t]
\caption{\label{tab:ikeda}Dimension vectors and NRMSE for the Ikeda map. Columns 2,3 and 4 contain the embedding
dimension vectors followed by their respective frequency of occurrence}
\begin{center}
{\small
\begin{tabular}{|l|l|l|l|l|l|l|}
\hline
     & \multicolumn{3}{c|}{Embedding dimensions} & \multicolumn{3}{c|}{NRMSE}\\
\hline
$N$    & FNN1    & FNN2   & PEM    & FNN1   & FNN2   & PEM    \\
\hline
512  & (1,1) 100 & (1,1) 100 & (2,2) 81     & 0.051 & 0.051 & 0.032 \\
     &             &             & (1,2) 13     &        &        &        \\
\hline
2048 & (1,1) 100 & (1,1) 100 & (2,2) 100    & 0.028 & 0.028 & 0.009 \\
\hline
8192 & (1,1) 100 & (1,1) 100 & (2,2) 100    & 0.013 & 0.013 & 0.003 \\
\hline
\end{tabular}
}
\end{center}
\end{table}

\begin{table}[t]
\caption{\label{tab:ikeda2}Dimension vectors and NRMSE for the
sum of two Ikeda maps}
\begin{center}
\begin{tabular}{|l|l|l|l|l|l|l|}
\hline
     & \multicolumn{3}{c|}{Embedding dimensions} & \multicolumn{3}{c|}{NRMSE}\\
\hline
$N$    & FNN1        & FNN2        & PEM            & FNN1   & FNN2   & PEM    \\
\hline
512  & (2,2) 89  & (2,2) 65  & (2,2) 63     & 0.456 & 0.480 & 0.447 \\
     & (3,3) 11  & (1,3) 26  & (1,2) 34     &        &        &        \\
\hline
2048 & (3,3) 95  & (2,3) 43  & (2,3) 54     & 0.339 & 0.365 & 0.329 \\
     & (2,2) 3   & (3,2) 24  & (2,2) 44     &        &        &        \\
\hline
8192 & (3,3) 100 & (2,3) 43  & (2,3) 100    & 0.260 & 0.304 & 0.251 \\
     &             & (1,4) 37  &                &        &        &        \\
\hline
\end{tabular}
\end{center}
\end{table}

\begin{table*}[t]
\caption{\label{tab:KDR}Dimension vectors and NRMSE for the KDR map}
\begin{center}
\begin{tabular}{|l|l|l|l|l|l|l|}
\hline
     & \multicolumn{3}{c|}{Embedding dimensions}           & \multicolumn{3}{c|}{NRMSE}\\
\hline
$N$    & FNN1             & FNN2            &    PEM          & FNN1   & FNN2   & PEM     \\
\hline
512  & (1,1,1,1) 100  & (0,0,2,2) 30 (1,1,1,1) 16  & (0,1,0,1) 80  (0,1,1,1) 14 & 0.776  & 0.907 & 0.629   \\
\hline
2048 & (1,1,1,1) 55 (2,2,2,2) 39  & (1,1,1,1) 37 (1,0,1,2) 21 & (0,2,1,1) 79  (0,1,0,1) 13  & 0.636  & 0.659 & 0.486   \\
\hline
8192 & (2,2,2,2) 85  (1,1,1,1) 15 & (2,1,1,1) 40 (1,1,1,1) 14 & (0,2,1,1) 100  & 0.558  & 0.551 & 0.373   \\
\hline
\end{tabular}
\end{center}
\end{table*}

\begin{table*}[t]
\caption{\label{tab:DRHenon}Dimension vectors and NRMSE for system of Driver-Response Henon system}
\begin{center}
\begin{tabular}{|l|l|l|l|l|l|l|l|}
\hline
     &      & \multicolumn{3}{c|}{Embedding dimensions}                    & \multicolumn{3}{c|}{NRMSE}\\
\hline
$N$    &  $C$   & FNN1         & FNN2                   &    PEM                & FNN1   & FNN2  & PEM      \\
\hline
512  &  0   & (2,2) 100  & (2,2) 98 (1,2)  1  & (2,2) 75 (2,1) 10 & 0.190  & 0.196 & 0.198    \\
\cline{2-8}
     &  0.4 & (2,2) 100  & (1,2) 89 (2,2)  8  & (3,2) 33 (2,2) 25 & 0.102  & 0.127 & 0.116    \\
\cline{2-8}
     &  0.8 & (2,2) 100  & (2,0) 99 (2,1)  1  & (3,0) 31 (0,3) 27 & 0.014  & 0.012 & 0.005    \\
\hline
2048 &  0   & (2,2) 100  & (2,2) 100            & (2,2) 100           & 0.093  & 0.093 & 0.093    \\
\cline{2-8}
     &  0.4 & (2,2) 100  & (1,2) 80 (2,2) 20  & (3,3) 45 (4,3) 45 & 0.050  & 0.084 & 0.028    \\
\cline{2-8}
     &  0.8 & (2,2) 100  & (2,0) 99 (2,1)  1  & (0,3) 20 (3,0) 19 & 0.007  & 0.006 & 0.001    \\
\hline
8192 &  0   & (2,2) 100  & (2,2) 100            & (2,2) 100           & 0.051  & 0.051 & 0.051    \\
\cline{2-8}
     &  0.4 & (2,2) 100  & (1,2) 100            & (3,3) 72 (4,3) 25 & 0.027  & 0.027 & 0.011    \\
\cline{2-8}
     &  0.8 & (2,2) 100  & (2,0) 98 (0,2)  2  & (0,4) 31 (4,0) 30 & 0.002  & 0.002 & 0.001    \\
\hline
\end{tabular}
\end{center}
\end{table*}

The FNN2 criterion generally gives multiple optimal $\mathbf{m}$ structures across realizations and PEM does the same
but only for small N. This indicates that high complexity degrades the performance of the algorithms for small sample
sizes. PEM is again best for predictions but overall we do not observe large differences in the three methods.

An interesting observation is that although FNN2 finds two optimal $\mathbf{m}$ with high frequencies they both give
the same $M$. This reflects the problem of identification, where different $\mathbf{m}$ unfold the attractor equally
well. This feature cannot be observed in FNN1 because the FNN1 algorithm inspects fewer possible vectors and only one
for each $M$, where $M$ can only be multiple of $p$ (in this case $(1,1)$ for $M$=2, $(2,2)$ for $M$=4, etc). On the
other hand, PEM criterion seems to converge to a single $\mathbf{m}$ for large $N$, which means that for the sum of the
two Ikeda maps this particular structure gives best prediction results. Note that there is no reason that the embedding
dimension vectors derived from FNN2 and PEM should match as they are selected under different conditions. Moreover, it
is expected that the $\mathbf{m}$ selected by PEM gives always the lowest average of multivariate NRMSE as it is
selected to optimize prediction.

\subsection{Kicked Double Rotor map}

The Kicked Double Rotor (KDR) map is a nonlinear chaotic system in four variables that describes the time evolution of
the mechanical system with the same name \cite{Grebgogi87}. The four time series $(x_{1,n},x_{2,n},y_{1,n},y_{2,n})$
are generated from the equations
\[
\mathbf{X}_{n+1}=\mathbf{M}\mathbf{Y}_n+\mathbf{X}_n, \quad
\mathbf{Y}_{n+1}=\mathbf{L}\mathbf{Y}_n+\mathbf{W}(\mathbf{X}_n)
\]
where
\begin{eqnarray}
\mathbf{X}_n=(x_{1,n},x_{2,n})^\top \in S^1\times S^1, \nonumber \\
\mathbf{Y}_n=(y_{1,n},y_{2,n})^\top \in \mathbb{R}\times
\mathbb{R}, \nonumber \\
\mathbf{W}(\mathbf{X}_n)=\bigl(6.36\sin(x_{1,n}),9\sin(x_{2,n})\bigr)^\top, \nonumber \\
\mathbf{M}=\left({0.49\atop 0.21} {0.21\atop 0.70} \right)
\quad \mathbf{L}=\left( {0.24\atop 0.27} {0.27\atop 0.51}
\right) \nonumber
\end{eqnarray}

The simulation results on four time series of KDR map in Table \ref{tab:KDR} are similar to those on the sum of two
Ikeda maps.

The selection of $\mathbf{m}$ from PEM outperforms the two FNN methods with respect to prediction and converges to a
single optimal $\mathbf{m}$ with $N$. For smaller $N$ there seems to be large diversity of selected $\mathbf{m}$ by all
methods.

\begin{table*}[t]
\caption{\label{tab:LHenon3}Dimension vectors and NRMSE for Lattice of 3 coupled Henon maps}
\begin{center}
\begin{tabular}{|l|l|l|l|l|l|l|l|}
\hline
     &      & \multicolumn{3}{c|}{Embedding dimensions}      & \multicolumn{3}{c|}{NRMSE}\\
\hline
$N$    &  $C$   & FNN1           & FNN2          &    PEM         & FNN1   & FNN2  & PEM      \\
\hline
512  &  0.4 & (2,2,2) 94 (1,1,1)  6  & (2,1,1) 46 (1,1,2) 37 & (1,2,1) 29 (1,1,2) 23   & 0.342  & 0.298 & 0.283    \\
\cline{2-8}
     &  0.8 & (2,2,2) 98 (1,1,1)  2  & (2,0,2) 91 (2,1,1)  4 & (2,0,2) 44 (2,1,1) 22  & 0.294  & 0.228 & 0.210    \\
\hline
2048 &  0.4 & (2,2,2) 100  & (1,2,1) 85 (1,1,2)  8 & (1,2,2) 34 (2,2,1) 30  & 0.169  & 0.203 & 0.170    \\
\cline{2-8}
     &  0.8 & (2,2,2) 100  & (2,0,2) 65 (1,2,1) 14 & (2,1,2) 48  (2,0,2) 41 & 0.119  & 0.131 & 0.112    \\
\hline
8192 &  0.4 & (2,2,2) 100  & (1,2,1) 100 & (2,2,2) 97 (3,2,3)  3  & 0.107  & 0.174 & 0.106    \\
\cline{2-8}
     &  0.8 & (2,2,2) 100  & (2,0,2) 100 & (2,1,2) 79 (3,2,3) 19  & 0.071  & 0.084 & 0.064    \\
\hline
\end{tabular}
\end{center}
\end{table*}

\begin{table*}[t]
\caption{\label{tab:LHenon4}Dimension vectors and NRMSE for Lattice of 4 coupled Henon maps}
\begin{center}
\begin{tabular}{|l|l|l|l|l|l|l|l|}
\hline
     &     & \multicolumn{3}{c|}{Embedding dimensions}           & \multicolumn{3}{c|}{NRMSE}\\
\hline
$N$    &  $C$   & FNN1             & FNN2            &    PEM          & FNN1   & FNN2   & PEM    \\
\hline
512  &  0.4 & (1,1,1,1) 100  & (1,1,1,1) 42 (1,0,2,1) 17 & (1,1,1,1) 45 (1,2,1,1) 20  & 0.285  & 0.363 & 0.288  \\
\cline{2-8}
     &  0.8 & (1,1,1,1) 100  & (1,1,1,1) 40 (1,0,1,2) 17 & (1,1,2,1) 25 (1,2,1,1) 17  & 0.314  & 0.357 & 0.291  \\
\hline
2048 &  0.4 & (1,1,1,1) 88 (2,2,2,2) 12   & (1,1,1,1) 88 (1,1,1,2)  7 & (1,2,2,1) 31 (2,1,2,1) 19  & 0.229  & 0.228 & 0.190  \\
\cline{2-8}
     &  0.8 & (1,1,1,1) 72 (2,2,2,2) 28  & (1,1,1,1) 36 (1,0,2,1) 33 & (2,1,1,2) 27 (2,2,1,1) 23  & 0.225  & 0.261 & 0.163  \\
\hline
8192 &  0.4 & (1,1,1,1) 85 (2,2,2,2) 15  & (1,1,1,1) 85 (1,2,1,1)  8 & (1,2,1,2) 46 (2,1,2,1) 45 & 0.197  & 0.200 & 0.137  \\
\cline{2-8}
     &  0.8 & (2,2,2,2) 86 (1,1,1,1) 14  & (1,2,0,1) 31 (1,0,2,1) 22 & (3,2,3,3) 79 (2,1,2,2) 13  & 0.131  & 0.209 & 0.072  \\
\hline
\end{tabular}
\end{center}
\end{table*}

\subsection{Driver-Response Henon system}

The Driver-Response Henon system consists of two Henon maps where the first Henon map (the variables $x_{1,n}$ and
$y_{1,n}$ are defined as in \eqref{eq:Henon}) drives the second Henon map \cite{Schiff96} as follows
\[
 \begin{array}{l}
  x_{2,n+1} = 1.4-(Cx_{1,n}x_{2,n}+(1-C)x_{2,n}^2)+y_{2,n} \\
  y_{2,n+1} = 0.3x_{2,n}
 \end{array}
\]
for a driving strength $C$. We set $C=0,0.4,0.8$ regarding three different states for the two systems: independent
($C$=0), moderately dependent ($C$=0.4) and strongly dependent ($C$=0.8). The results of the Monte Carlo simulations
for the bivariate time series of $(x_{1,n},x_{2,n})$ are given in Table \ref{tab:DRHenon}.

First we observe that FNN1 gives uniform results for all $C$ and $N$. When the two time series are independent ($C$=0)
there is actually no reason to apply multivariate embedding and all methods select for all $N$ the same embedding
dimension vector (2,2) (less than 100\% frequency only for $N$=512 with FNN2 and PEM). Since the optimal embedding
dimension for the Henon map is known to be 2 this result seems quite reasonable.

When $C$=0.4 the moderate dependence of the second time series to the first affects the selection of the embedding
dimension vector. FNN2 selects mostly $\mathbf{m}=(1,2)$, which means that this method detects that information of the
driver time series is passed to the response, thus it utilizes more the response time series for unfolding the
attractor. On the other hand PEM tends to select vectors with larger embedding dimension for the driver ((3,2) for
$N$=512 and (4,3) for $N$=2048) because this information is more useful for prediction purposes. Also PEM gives
over-embedding as for the sum of two Ikeda maps.

The strong dependence of the second time series to the first when $C$=0.8 implies that the system is less complex and
so a smaller $M$ is needed for embedding. For all $N$, FNN2 almost always selects the vector (2,0), whereas PEM cannot
distinguish the two time series and selects with almost equal frequencies vectors of the form $(m,0)$ and $(0,m)$
giving again over-embedding as $N$ increases. Thus PEM does not reveal the coupling structure of the underlying system
and picks any embedding dimension structure among a range of structures that give essentially equivalent predictions.
Here FNN2 seems to detect sufficiently the underlying coupling structure in the system resulting in a smaller total
embedding dimension that gives however the same level of prediction as the larger $M$ suggested by FNN1 and slightly
smaller than the even larger $M$ found by PEM.

\subsection{Lattices of coupled Henon maps}

The last system is an example of spatiotemporal chaos and is defined as a lattice of $k$ coupled Henon maps
$\{x_{i,n},y_{i,n}\}_{i=1}^{k}$ \cite{Politi92} specified by the equations
\[
 \begin{array}{l}
 x_{i,n+1}=1.4-((1-C)x_{i,n}+ {C(x_{i-1,n}+x_{i+1,n}) \over
 2})^2+y_{i,n} \\
 y_{i,n+1}=0.3x_{i,n}
 \end{array}
\]
The connection of the $k$ maps is restricted between adjacent maps in the ordered list of maps, i.e. each $x_{i,n},\
i=2,...,k-1$ is connected to $x_{i+1,n}$ and $x_{i-1,n}$, with the ``boundary'' maps for $i=1,k$ being simple Henon
maps. As in the case of driven-response Henon maps the complexity of the system and the nature of the dependence of the
time series with each other is determined by their coupling strength, which here is fixed for all couplings in the
lattice. The results for coupling strengths C=$0.4$ and $0.8$ and two lattice structures for $k$=3 and $k$=4 are given
in Tables \ref{tab:LHenon3} and \ref{tab:LHenon4}, respectively.

For both lattices the results are similar to that of the driven-response Henon maps. For $k=3$, PEM does not single out
an $\mathbf{m}$ structure for small sample sizes or for moderate coupling, where FNN2 generally does. PEM again gives
the best prediction results and FNN2 is more conservative giving always the smallest $M$ of all three methods. The
lattice involving 4 maps is a more complicated system since there are more possible embedding dimension vectors for a
given $M$ and thus there is more diversity in the results. We note that for almost all combinations of data size,
coupling strength and method (meaning FNN2 and PEM) there are multiple selected optimum dimension vectors. Beyond this,
the differences in FNN2 and PEM discussed for $k=3$ persist also for $k=4$.

\section{Discussion}
\label{sec:Discussion}

There does not seem to be an optimal scheme for state space reconstruction from multivariate time series. The
simulation results on two schemes proposed in this work, one based on unfolding the attractor at any possible direction
(FNN2) and the other aiming at optimizing prediction performance (PEM), seem to confirm this.

When the goal of state space reconstruction is to make predictions, selection of multivariate embedding with the
prediction criterion PEM is best, but this results often to over-embedding (large total embedding dimension) and does
not really estimates the actual degrees of freedom of the underlying system. This can also be justified from the fact
that the dimension of the reconstructed state space selected by PEM tends to increase with the sample size, at least
for the sizes we used in the simulations. Such a feature shows lack of consistency of the PEM criterion and suggests
that the selection is led from factors inherent in the prediction process rather than the quality of the reconstructed
attractor. For example the increase of embedding dimension with the sample size can be explained by the fact that more
data lead to abundance of close neighbors used in local prediction models and this in turn suggests that augmenting the
embedding vectors would allow to locate the $K$ neighbors used in the model.

On the other hand, the two schemes used here that extend the method of false nearest neighbors (FNN) to multivariate
time series aim at finding minimum embedding that unfolds the attractor, but often a higher embedding gives better
prediction results. In particular, the second scheme (FNN2) that explores all possible embedding structures gives
consistent selection of an embedding of smaller dimension than that selected by PEM. Moreover, this embedding could be
justified by the underlying dynamics of the known systems we tested. However, lack of consistency of the selected
embedding was observed with all methods for small sample sizes (somehow expected due to large variance of any estimate)
and for the coupled maps (probably due to the presence of more than one optimal embeddings).

In this work, we used only a prediction performance criterion to assess the quality of state space reconstruction,
mainly because it has the most practical relevance. There is no reason to expect that PEM would be found best if the
assessment was done using another criterion not based on prediction. However, the reference (true) value of other
measures, such as the correlation dimension, are not known for all systems used in this study. Another constraint of
this work is that only noise-free multivariate time series from discrete systems are encountered, so that the delay
parameter is not involved in the state space reconstruction and the effect of noise is not studied. It is expected that
the addition of noise would perplex further the process of selecting optimal embedding dimension and degrade the
performance of the algorithms. For example, we found that in the case of the Henon map the addition of noise of equal
magnitude to the two time series of the system makes the criteria to select any of the three equivalent embeddings
($(2,0)$,$(0,2)$,$(1,1)$) at random. It is in the purpose of the authors to extent this work and include noisy
multivariate time series, also from flows, and search for other measures to assess the performance of the embedding
selection methods.

\begin{acknowledgments}
This paper is part of the 03ED748 research project, implemented within the framework of the "Reinforcement Programme of
Human Research Manpower" (PENED) and co-financed at 90\% by National and Community Funds (25\% from the Greek Ministry
of Development-General Secretariat of Research and Technology and 75\% from E.U.-European Social Fund) and at 10\% by
Rikshospitalet, Norway.
\end{acknowledgments}

\bibliography{ivla2}

\end{document}